# A SIMPLE PHYSICAL INTERPRETATION OF THE LENSE-THIRRING EFFECT BASED ON EMQG THEORY

Tom Ostoma and Mike Trushyk
Email: emgg@rogerswave.ca
Wednesday, March 16, 1999

## ABSTRACT

We use ElectroMagnetic Quantum Gravity (EMQG, ref. 1) to provide a simple physical model of the Lense-Thirring effect on the earth. The Lense-Thirring effect is a tiny perturbation of the motion of a free-falling particle near a massive rotating object, first calculated by the physicists J. Lense and H. Thirring in 1918 using general relativity. The Lense-Thirring effect can also be thought of as the 'dragging of inertial frames', as first named by Einstein himself. The amount of frame dragging varies with height above the spinning earth. Einstein's general relativity predicts the value of this effect near the earth, but the effect has not yet been accurately verified experimentally. However, recent work using the LAGEOS series, earth orbiting satellites has rendered an unconfirmed experimental value that agrees with theory to an accuracy of about 20 %. Our physical model is based on the disturbance of the electrically charged virtual particles of the quantum vacuum in a state of free-fall near a massive spinning object. We calculate the Lense-Thirring effect on the earth's surface using this model that is in good agreement with the value derived from general relativity.

Inertial frame dragging for an observer on a large, rapidly spinning mass results from the following:

(1) Gravitons originating from a rapidly spinning, massive object move outwards at a finite speed, which is the velocity of light. This allows time for the spinning mass to rotate and carry an observer along by a small amount while the gravitons are still in flight.
(2) As the gravitons propagate outward, they encounter virtual, electrically charged fermion particles in the quantum vacuum. Since all fermions posses 'mass-charge', the quantum vacuum is accelerating downwards in the opposite direction of the motion of the graviton particle flux, with the magnitude of the acceleration depending on the graviton flux at that point.
(3) Since in EMQG, inertia results from the <u>relative acceleration</u> of the electrically charged particles that make up a test mass, with respect to the electrically charged virtual fermion particles of the quantum vacuum, the free-fall direction of a test mass is the same as the virtual particle free-fall.
(4) Gravitons are physically very similar to photons in EMQG, and the graviton flux can be visualized as being the same as the photon flux from a rotating flashlight. Since the velocity of the source does not effect the velocity of photons, the velocity of the spinning earth does not affect the velocity of the gravitons (in the weak field limit).
(5) Thus, from the perspective of an <u>external</u> observer, the mass falls along the straight radius vectors. The observer on the <u>surface of the equator</u> 'sees' the graviton flux leaving the equator in a curved, spiral path, as he is carried away with the rotation. From his frame of reference the graviton path appears to curve, and the average acceleration vector of the virtual fermion particles follows this same curved path, with the magnitude of the virtual fermion acceleration vectors increasing in magnitude the closer to the spinning mass.

The finite velocity of the graviton particle, along with the apparent shift of the downward acceleration of the electrically charged virtual fermion particles of the quantum vacuum is entirely responsible for inertial frame dragging. Therefore, the natural 'force-free' path that a free-falling test mass will follow appears curved (or dragged) by the earth's rotation. These curved paths also represent the path that light will take if it moves 'straight' up, or in other words, a geodesic path.

## 1. INTRODUCTION TO THE LENSE-THIRRING EFFECT

The Lense-Thirring effect is a tiny perturbation of the orbit of a particle caused by the spin of the attracting body, first calculated by the physicists J. Lense and H. Thirring in 1918 using general relativity (ref. 8). Einstein's general relativity predicts the magnitude of this perturbation in the vicinity of the spinning body, but the effect has not been verified experimentally. However, recent work (ref. 9) using the LAGEOS and LAGEOS II earth orbiting satellites has rendered an unconfirmed experimental value that agrees with general relativity theory to an accuracy of about 20 %. The Lense-Thirring effect has also be interpreted as due to gravitomagnetic fields (section 7), and also ties in with Mach's principle (ref. 10). It is hoped that with the launch of the Gravity Probe B (co-developed by Stanford University) by NASA the Lense-Thirring effect will be measured to an unprecedented accuracy of 1% or better. The Gravity Probe B (there was a different Gravity Probe A launched earlier by NASA) is a drag free satellite carrying four ultra-precise gyroscopes that will be put in a polar orbit around the earth at a height of about 400 miles (ref. 11).

An important consequence of the Lense-Thirring effect is that the orbital period of a test mass around the earth depends on the direction of the orbit! A test mass that has an orbit which revolves around the earth in the same direction as the spin rotation would be longer then the orbital period of the same test mass revolving opposite to the direction of the spin of the earth. The difference in the orbital period of the two test masses becomes smaller with increasing height until it disappears when the orbits are at infinity.

The Lense-Thirring effect can be thought of as a kind of 'a dragging of inertial frames', first named by Einstein himself. What this means is that the free fall (gravity-free) motion of a test mass is modified in the presence of a large massive spinning sphere, as compared to the identical case of the non-rotating sphere. If a test mass is in free-fall, the contribution of gravity to the motion of the mass is canceled out (at a point), and the resulting motion of the test mass is purely inertial in nature. In other words, free fall masses ought to take on motions that are the same as that in far space, away from gravitational sources. However, the inertial motion of the mass near the earth is different due to the presence of the Lense-Thirring effect, as compared to a non-rotating earth. As a result of the Lense-Thirring effect, the earth seems to 'drag the local inertial reference frame'. Therefore the small orbiting gyroscopes in Gravity Probe B in earth polar orbit with the spin axes oriented towards a distant star will have is its spin axis changed by the Lense-Thirring effect as it orbits.

Unfortunately the general relativistic approach to the Lense-Thirring effect is very mathematical in nature, and the mathematics currently defies any simple physical interpretation of the phenomena. In fact, it is very hard to see how the spin of a massive object can drag the very fabric of 4D space-time. In general relativity, the Lense-Thirring calculation is based on the weak field approximation of Einstein's famous gravitational field equations.

## 2. INTRODUCTION TO EMQG AND INERTIAL FRAME DRAGGING

ElectroMagnetic Quantum Gravity or EMQG (ref. 1) was developed to provide a quantum model of gravity that is manifestly compatible with a Cellular Automata model of the universe (ref. 2 and 4). EMQG is also a theory that is based totally on the particle exchange paradigm for *all* forces, where gravity is no exception. The particle exchange paradigm fits very well within the Cellular Automata framework. EMQG is also in accordance with the general principles of quantum field theory.

Cellular Automata (CA) is a computer model that consists of a large number of cells, which are storage locations for numbers. Each cell contains some initial number (say 0 or 1 for example), and the same set of rules are applied for *each* and every cell. The rules specify how these numbers are to be changed at the next computer 'clock' interval. Mathematically, a 'clock' is required in order to synchronize the next state of all the cells. The logical rules of a cell specifies the new state of that cell on the next 'clock' period, based on the cells current state, and on that of all the cell's immediate neighbors (each cell in CA has a fixed number of neighboring cells). The number of neighbors that influence a given cell is what we call the connectivity of the cellular automata. In other words, the number of neighbors that connect (or influence) a given cell is called the CA connectivity. The connectivity can be any positive integer number. EMQG was the result of intensive investigations into a model for gravity that fits the general principles of CA theory and the particle exchange paradigm. Edward Fredkin has been the first person credited with introducing the idea that our universe is a huge cellular automata computer simulation.

When developing EMQG, we looked for a mechanism that produces the gravitational force, which is somehow linked to the Einstein principle of equivalence of inertial and gravitational mass. In addition, this mechanism should naturally lead to 4D space-time curvature and should also be compatible with the principles of general relativity theory. Nature has another long-range force called electromagnetism, which has been described successfully by the principles of quantum field theory. This well known theory is called Quantum ElectroDynamics (QED), and this theory has been tested for electromagnetic phenomena to an unprecedented accuracy. It is therefore reasonable to assume that gravitational force should be a similar process, since gravitation is also a long-range force like electromagnetism. However, a few obstacles lie in the way that complicate this line of reasoning:

(1) **Gravitational force is observed to be always attractive**! In QED, electrical forces are attractive and repulsive. There are equal number of positive and negative charged virtual particles in the quantum vacuum at any given time because virtual particles are always created in equal and opposite, electrically charged pairs in accordance with the conservation of electric charge. Thus there is always a balance of attractive and repulsive electrical forces in the quantum vacuum, and the quantum vacuum is essentially electrically neutral, overall. If this were not the case, the virtual charged particles of one electrical charge type in the vacuum would dominate over all other physical processes. If the vacuum contained only positively charged virtual particles, there would essentially be a huge cosmological constant problem from the perspective of electrical charge. However, according to general relativity gravity and the principle of equivalence, gravitational force is always attractive. This statement also includes anti-matter. However, if gravity is interpreted as a 'mass charge' in accordance with quantum field theory, than there is no cancellation of the mass charge for the virtual particles of the vacuum. Thus, we are left with the problem of how so much mass in the quantum vacuum amounts to almost no effect on the overall state of the universe.

(2) **QED is formulated in a relativistic, flat 4D space-time with no curvature**. In QED, electrical charge is defined as the fixed rate of emission of photons (strictly speaking, the fixed probability of emitting a photon) from a given charged particle. Electrical forces are caused by the exchange of photons, which propagate between the charged particles. The photons transfer momentum from the source charge to the destination charge, and travel in flat 4D space-time (assuming no background gravity). From these basic considerations, a successful theory of quantum gravity should have an exchange particle (graviton), which is emitted from a mass particle at a fixed rate as in QED. The 'mass charge' replaces the idea of electrical charge in QED, and the graviton momentum exchanges are now the root cause of gravitational force. Yet, the graviton exchanges must somehow produce disturbances of the normally flat space and time, when originating from a very large mass.

It is strange that gravity, which is also a long-range force, is governed by same form of mathematical law as found in Coulomb's Electrical Force law. Coulomb's Electric Force law states: $F = KQ_1Q_2/R^2$, and Newton's Gravitational Force law: $F=GM_1M_2/R^2$. Therefore, one would expect that these two forces should be very similar. This suggests that there is a deep connection between gravity and electromagnetism. Yet, gravity supposedly has no counterpart to negative electrical charge. Thus, there seems to be no such thing as negative 'mass charge' for gravity, as we find for electrical charge. Furthermore, QED also has no analogous phenomena as the principle of equivalence. Why should gravity be connected with the principle of equivalence, and thus inertia, and yet no analogy of this exists for electromagnetic phenomena?

In response to the question of negative 'mass charge', *EMQG postulates the existence of negative 'mass charge' for gravity, in close analogy to electromagnetism*. Furthermore, **we claim that negative 'mass charge' is possessed by all anti-particles that carry mass**. Therefore anti-particles, which are opposite in electrical charge to ordinary particles, are also opposite in 'mass charge'. In fact, negative 'mass charge' is not only abundant in nature, it comprises nearly half of all the mass particles (in the form of 'virtual' particles) in the universe! The other half exists as positive 'mass charge', also in the form of virtual particles. Furthermore, all familiar ordinary (real) matter comprises only a vanishing small fraction of the total 'mass charge' in the universe! Real anti-matter seems to be very scarce in nature, and no search for it in the cosmos has revealed it's existence in bulk form to date.

Both positive and negative 'mass charge' appear in huge numbers in the form of virtual particles, which quickly pop in and out of existence in the quantum vacuum, everywhere in empty space. *The existence of negative 'mass charge' is the key to the solution to the famous problem of the cosmological constant* (reference 1), which is one of the great unresolved mysteries of modern physics. **Finally, we propose that the negative energy or the antimatter solution of the famous Dirac equation of quantum field theory is also the negative 'mass charge' solution of that equation.**

With EMQG, we discovered the hidden quantum processes behind the Einstein Equivalence principle (ref. 1) and the quantum origins of Newtonian Inertia (ref. 3). We also discovered a tiny flaw in the Einstein Weak Equivalence Principle (WEP, ref. 1), which makes the WEP invalid in general. It also turns out the virtual particles of the quantum vacuum plays an *essential role in gravity*, and also for *inertia*. In EMQG, the virtual particles of the quantum vacuum act like a reference for accelerated motion (but *not* for constant velocity motion). The problems and paradoxes introduced in Mach's principle are easily resolved within this framework. Therefore, any discussion of inertial reference frames and inertial dragging must take into consideration the state of the background, electrically charged virtual particles of the vacuum.

Here we will show that the Lense-Thirring effect is really due to the dynamics of the electrically charged, virtual particles of the quantum vacuum, which is affected by a massive rotating object through the process of graviton exchanges. The interaction of the vacuum with the charged particles inside a test mass accounts for the dragging of the inertial frame of that test mass. It turns out that the interaction of the electrically charged, virtual particles of the quantum vacuum with the electrically charged particles that make up a mass is also ***the actual source of inertial mass***. We call this new theory of inertia 'Quantum Inertia' or QI. Since inertial frames play a central role in the Lense-Thirring effect, we provide a brief review of QI in the next section. A more detailed account of quantum inertia can be found in reference 1.

### 3.      INTRODUCTION TO QUANTUM INERTIA THEORY

**"Under the hypothesis that ordinary matter is ultimately made of subelementary constitutive primary charged entities or 'partons' bound in the manner of traditional elementary Plank oscillators, it is shown that a heretofore uninvestigated Lorentz force (specifically, the magnetic component of the Lorentz force) arises in any accelerated reference frame from the interaction of the partons with the vacuum electromagnetic zero-point-field (ZPF). ... The Lorentz force, though originating at the subelementary parton level, appears to produce an opposition to the acceleration of material objects at a macroscopic level having the correct characteristics to account for the property of inertia."**

                                                                                          **- B. Haisch, A. Rueda, H. E. Puthoff**

Inertia is one of the most important concepts in theoretical physics. It forms the core concept in Newton's laws of motion. Yet our basic understanding of inertia has changed little since Newton first conceived his famous law of inertia: F = ma. Quantum mechanics teaches us that mass is really an ensemble of quantum particles interacting through forces.

Therefore, there ought to be some quantum explanation as to why quantum particles that make up a mass opposes acceleration through an inertial force.

To this end, it has been recently proposed that Newtonian Inertia is strictly a quantum vacuum phenomena, where inertial force is the ***sum of all the electrical force interactions*** that exists between the charged particles of a mass and the surrounding virtual particles of the quantum vacuum! If this is true, then the existence of the quantum vacuum actually reveals it's presence to us in almost all daily activities. Unlike the hard-to-measure quantum vacuum effect called the Casimir effect (section 4), the presence of the inertial force is obvious and universal. The consequences of inertia prevail throughout all of physics. For example, the orbital motion of the earth around the sun is actually a balancing act between two forces: inertia and gravitation.

R. Haisch, A. Rueda, and H. Puthoff (in 1994) were the first to propose a theory of inertia on the quantum scale (known here as HRP Inertia, reference 5), where the quantum vacuum plays a central role in Newtonian inertia. They suggested that inertia is due to the strictly local electrical force interactions of charged particles that make up a mass with the immediate background virtual particles of the quantum vacuum (in particular the virtual photons or ZPF as the authors called it). They found that inertia is caused by the magnetic component of the Lorentz force, which arises between what the author's call the charged 'parton' particles in an accelerated reference frame interacting with the background quantum vacuum virtual particles. The sum of all these tiny forces in this process is the source of the resistance force opposing accelerated motion in Newton's F=MA. The 'parton' is a term that Richard Feynman coined for the constituents of the nuclear particles such as the proton and neutron (now called quarks).

We have built on their work and found that it is necessary to make a small modification to HRP Inertia theory as a result of our investigation into the principle of equivalence. Our modified version of HRP inertia is called "Quantum Inertia" (or QI), and is described in detail in reference 1. This theory also resolves the long outstanding problems and paradoxes of accelerated motion introduced by Mach's principle, by suggesting that the vacuum particles themselves serve as Mach's universal reference frame (for <u>acceleration</u> only), without violating the principle of relativity of constant velocity motion. In other words, our universe offers no observable reference frame to gauge inertial frames (non-accelerated frames where Newton's laws of inertia are valid), because the quantum vacuum offers no means to determine absolute constant velocity motion. However for accelerated motion, the quantum vacuum plays a very important role by offering a resistance to acceleration, which results in an inertial force opposing the acceleration of the mass. Thus, the very existence of inertial force reveals the absolute value of the *acceleration* with respect to the net statistical average acceleration of the virtual particles of the quantum vacuum.

In the past various clues have surfaced as to the importance of the state of the virtual particles of the quantum vacuum to accelerated motion of real charged particles. One example is the so-called Davies-Unruh effect (ref. 15), where uniform and linearly accelerated charged particles in the vacuum are immersed in a heat bath, with a temperature proportional to acceleration (with the scale of the quantum heat effects being very low). However, the work of reference 5 is the first place we have clearly seen the identification of inertial forces as the direct consequence of the interactions of real matter particles with the virtual particles of the quantum vacuum.

In HRP Inertia, Haisch, Rueda, and Puthoff have come up with a new version of Newton's second law: F=MA. As in Newton's theory, their expression has 'F' for force on the left-hand side and 'A' for acceleration on the right. But in the place of 'M', there is a complex mathematical expression tying inertia to the properties of the vacuum. They found that the fluctuations in the vacuum interacting with the charge particles of matter in an accelerating mass give rise to a magnetic field, and this in turn, creates an opposing force to the motion. Thus, electromagnetic forces (or photon exchanges) is ultimately responsible for the force of inertia! The more massive an object, the more 'partons' it contains; and the more partons a mass contains means more individual (small) electromagnetic forces from the vacuum are present and the stronger the reluctance to undergo acceleration. But, when a mass is moving at a **constant** velocity, inertia disappears, and there is no resistance to motion in any preferred direction, as required by special relativity.

In HRP version of inertia theory, inertia is caused by the magnetic component of the Lorentz force which arises between what the author's call 'parton' particles in an accelerated reference frame interacting with the background vacuum electromagnetic zero-point-field (ZPF). The author's use the old fashion term originated by Feynman called the 'parton', which referred to the elementary constituents of the nuclear particles such as protons and neutrons. It is now known that Feynman's partons are quarks, and that the proton and neutron each contain three quarks of two types: called the 'up' and 'down' quarks.

We have found it necessary to make a small modification of HRP Inertia theory in our investigation of the principle of equivalence. In EMQG, the modified version of inertia is known here as the "Quantum Inertia", or QI. In EMQG, a new elementary particle is required to fully understand inertia, gravitation, and the principle of equivalence. **All** matter, including electrons and quarks, must be made of nature's most fundamental mass unit or particle which we call the 'masseon' particle. These particles contain one fixed, fundamental 'quanta' of both inertial and gravitational mass. The masseons also carry one basic, smallest unit or quanta of electrical charge as well, of which they can be either positive or negative. Masseons exist in particle or anti-particle form (called anti-masseon), that can appear at random in the vacuum as masseon/anti-masseon particle pairs of opposite electric charge. The earth consists of ordinary masseons (no anti-masseons), of which there are equal numbers of positive and negative electric charge varieties. The masseon particle model will be elaborated later. Instead of the 'parton' particles (that make up an inertial mass in an accelerated reference frame) interacting with the background vacuum electromagnetic zero-point-field (ZPF), we postulate that the real masseons (that make up an accelerated mass) interacts with the surrounding, virtual masseons of the quantum vacuum, electrically. However, the detailed nature of this interaction is still not known at this time, and therefore Quantum Inertia remains as a postulate of EMQG theory.

We found that quantum inertia is deeply connected with the subject of quantum gravity. EMQG explains why the inertial mass and gravitational mass are identical in accordance with the weak equivalence principle. The weak equivalence principle translates to the simple fact that the mass (m) that measures the ability of an object to produce (or react to) a gravitational field (F=GMm/r$^2$) is the same as the inertial mass value that appears in Newton's F=ma. In EMQG, this is not a chance coincidence, or a given fact of nature, which is assumed to exist, *a prior*. Instead, equivalence follows from a deeper process occurring inside a gravitational mass due to interactions with the quantum vacuum, which are *very similar* in nature to the interactions involved in inertial mass undergoing acceleration. Since the state of motion of the virtual particles of the quantum vacuum is central to our model of the Lense-Thirring effect, we briefly review the quantum vacuum from the perspective of EMQG.

## 4. THE VIRTUAL PARTICLES OF THE QUANTUM VACUUM

How does one visualize an absolutely perfect vacuum? Using classical physics, one might imagine that the vacuum is completely devoid of everything. According to qunatum theory, the vacuum is far from empty. In order to produce a perfect vacuum, one must remove all matter from an enclosure. However, this is still not good enough. One must also lower the temperature down to absolute zero in order to remove all thermal electromagnetic radiation present from the surrounding environment.

Nernst correctly deduced in 1916 (ref. 8) that empty space is still not completely devoid of all radiation after this is done. He predicted that the vacuum is still permanently filled with an electromagnetic field propagating at the speed of light, called the zero-point fluctuations (sometimes called vacuum fluctuations). This was later confirmed by the full quantum field theory developed in the 1920's and 30's. Later, with the development of QED, it was realized that all quantum fields should contribute to the vacuum state, like virtual electrons and positron particles, for example.

According to modern quantum field theory, the perfect vacuum is teeming with all kinds of activity, as all types of quantum virtual matter particles (and virtual bosons or force particles) from the various quantum fields, appear and disappear spontaneously. These particles are called 'virtual' particles because they result from quantum processes that have small energies and very short lifetimes, and are therefore undetectable.

One way to look at the existence of the quantum vacuum is to consider that quantum theory forbids the absence of motion, as well as the absence of propagating fields (exchange particles). This follows from the Heisenberg uncertainty principle. In QED, the quantum vacuum consists of the virtual particle pair creation/annihilation processes (for example, electron-positron pairs), and the zero-point-fluctuation (ZPF) of the electromagnetic field (virtual photons) just discussed. The existence of virtual particles of the quantum vacuum is essential to understanding the famous Casimir effect (ref. 9), an effect predicted theoretically by the Dutch scientist Hendrik Casimir in 1948. The Casimir effect refers to the tiny attractive force that occurs between two neutral metal plates suspended in a vacuum. He predicted theoretically that the force 'F' per unit area 'A' for plate separation D is given by:

F/A  =  - $\pi^2$ h c /(240 D$^4$ )    Newton's per square meter   (Casimir Force 'F')        (4.1)

The origin of this minute force can be traced to the disruption of the normal quantum vacuum virtual photon distribution between two nearby metallic plates. Certain photon wavelengths (and therefore energies) in the low wavelength range are not allowed between the plates, because these waves do not 'fit'. This creates a negative pressure due to the unequal energy distribution of virtual photons inside the plates as compared to outside the plate region. The pressure imbalance can be visualized as causing the two plates to be drawn together by radiation pressure. Note that even in the vacuum state, virtual photons carry energy and momentum.

Recently, Lamoreaux made (ref. 10) accurate measurements for the first time on the theoretical Casimir force existing between two gold-coated quartz surfaces that were spaced 0.75 micrometers apart. Lamoreaux found a force value of about 1 billionth of a Newton, agreeing with the Casimir theory to within an accuracy of about 5%.

5.        **EMQG AND THE MEANING OF INERTIAL FRAME DRAGGING**

The first step to solving any problem in EMQG theory is to determine the motion of the virtual particles of the quantum vacuum. To determine this, we need to know the graviton flux and direction at every point. First we must recall that according to EMQG, the graviton particle is very similar in nature to the photon particle. The motion of a light source does not affect the velocity of light. Similarly, the motion of the rotating earth does not affect the velocity of the graviton flux leaving the earth. The virtual particles of the quantum vacuum surrounding the earth absorb some of the graviton flux, which results in a downward acceleration in the opposite direction as the outward graviton flux. The magnitude of a virtual particle acceleration depends on the received graviton flux for that particle.

Since the graviton flux rate decreases with the inverse square of the distance from the earth (due to geometric spreading of the flux), the virtual particles accelerate downwards at an acceleration that decreases with height. If observer 'A' on the earth drops a mass, the mass takes on the same net downward acceleration as the virtual particles of the vacuum, through the quantum inertia mechanism described above. According to EMQG, the path of the accelerated virtual particles of the quantum vacuum dictates the direction and magnitude of the 4D space-time curvature (reference 1). Normally it is perpendicular to the surface, but some physical circumstances may alter this.

We can now study details of the Lense-Thirring effect for the rotating earth. The Lense-Thirring effect is more pronounced as the angular velocity of the earth is increased. The basic reason for inertial frame dragging is the finite speed of propagation of the graviton particle (which is the speed of light). This allows time for a large rapidly spinning mass to rotate a small amount while the gravitons are still in flight as it propagates outwards. ***The finite velocity of the graviton particle along with the downward $GM/R^2$ acceleration component of the charged virtual particles of the quantum vacuum is entirely responsible for inertial frame dragging***. This is the fundamental reason for inertial frame dragging.

In order to simplify our analysis, we will assume that most of the mass of the earth is concentrated in a small spherical region at the earth's core, with the rest of the earth's volume containing a negligible quantity of mass. The graviton moves outward at the speed of light from the spinning core (neglecting the virtual particle scattering effects, which reduces its absolute velocity (in CA units) as it propagates upwards from the earth's core. Gravitons are emitted in huge numbers from the earth's core. Since we know that gravitons are physically very similar to photons in EMQG, we can predict the characteristics of this graviton flux. The graviton flux can be visualized as being the same as a rotating flashlight emitting light outwards as it rotates. Since the velocity of the source does not effect the velocity of light, we conclude that the velocity of the spinning core does not affect the motion of the gravitons.

The inset in the upper right hand corner of figure 9 shows the view of the earth looking down at the north pole, where the earth is rapidly rotating at an angular velocity close to light velocity. The non-rotating observer 'B' is stationed above the equator, and 'sees' the graviton flux leaving the equator along the radius vectors. It is important to note that when the graviton leaves the rotating source, the motion of the graviton is totally de-coupled from the source. The gravitons move at a constant velocity along straight radius lines, and are not affected by the motion of the earth in any way. However, gravitons are affected by the state of accelerated motion of the virtual particles. The note at the end of this section gives more detail on this important point, and the reason why inertial frame dragging is highly non-linear for very massive objects rotating at relativistic speeds. For this analysis, we can neglect this effect for the earth's relatively slow rotation rate.

As the gravitons propagate outward they encounter virtual masseon particles in the quantum vacuum, and some of the gravitons are absorbed . Since virtual masseons posses 'mass-charge', these particles are accelerated downward, or fall. The virtual masseons are falling in the same direction as the source of the graviton particle flux. The magnitude of the acceleration depends on the graviton flux at a point 'r' from the center, and is given by $GM/r^2$. If a test mass (composed of real electrically charged masseon particles) is dropped near the surface of the earth, the electrical interactions between the real masseons of the mass and the virtual masseons causes the test mass to accelerate downwards at 1g in the direction of the graviton flux. Thus, from the perspective of observer 'B' off the earth's surface, the mass falls along the straight radius vectors. What does observer 'A' on the earth's surface see?

Figure 9 shows our view of the earth looking down at the north pole, where the earth is rapidly rotating at an angular speed close to light velocity to emphasis the effect. Observer 'A' is situated on the surface of the equator and 'sees' the graviton flux leaving the equator in curved paths. Why is the graviton flux curved? Recall that the graviton flux is moving in 'straight' line paths along the radius vectors. However, observer 'A' is carried along with the rotation of the earth. Therefore, from his frame of reference the gravitons appear to curve. The average acceleration vector of the virtual particles follows this curved path, with the acceleration vectors increasing in magnitude the closer to the center. A test mass dropped on the surface of the earth by observer 'A' moves along the curved path in figure 9, guided by the motion of the virtual particles by photon exchanges. These curved paths also represent the path that light will take if it propagates straight up (in other words, geodesic paths). In absolute CA units, the light velocity varies upwards along these curved paths from observer's A reference frame, due to the Fizeau-like scattering with the quantum vacuum (reference 1). According to EMQG, these paths represent the actual direction of the 4D space-time curvature in observer's 'A' frame of reference. These paths are deflected for observer 'A' compared to when the earth is not rotating.

The outward propagating graviton curve in the equatorial plain for a clockwise rotating earth turns out to be Archimedes' spiral, which takes the form $r = k\theta$ in polar coordinates, where k is a constant (r is the distance that the graviton travels). The constant k depends only on the velocity of the graviton 'c' and the angular velocity ($v = 2\pi R/T_p$) of the earth's rotation, where $T_p$ is the rotation period of the earth and R is the earth's radius, and the time t is the time of transit of the graviton. If k is small then the spiral has a high curvature, and if k is large the curvature is small. The ratio 'c/v' determines the value of the spiral constant k. If c were to be very large, then $k \gg 1$ which causes the gravitons to appear to 'unwind' slowly ; and if v were to be very large, then $k \ll 1$ which causes the gravitons to appear to 'unwind' rapidly. Note: this is all from observers 'A' frame.

Based on these considerations, it is easy to show that the equation for the spiral is:

$r = c^2 \theta / (2\pi v)$   where $v = 2\pi R/T_p$ , and $\theta = 2\pi t v/c$ (5.1)

The Archimedes spiral is normalized such that if v=c, $\theta = 2\pi$, and t=1 second then the value of r is 300,000 km after one rotation. To verify that eq. 5.1 is correct, let us see how r varies with $\theta$. When the velocity of rotation is v=c, then according to eq. 5.1 the equation for the spiral is $r = [c/(2\pi)]\theta$. When the graviton travels for one rotation of the observer A, the spiral path appears to twist once, with $\theta = 2\pi$ radians, and r = 300,000 km. Similarly, when the graviton travels for two rotations of the observer A, the spiral path appears to twist once, with $\theta = 4\pi$ radians, and r = 600,000 km. If v=c/2, then if t = 2 seconds, $\theta = 2\pi$, and r = 600,000 km.

Let us calculate the shape of the spiral for the observer 'A' on the earth. The earth has a radius $R = 6.37 \times 10^6$ meters, and a rotation period $T_p = 24$ hours. Therefore, the v = 463 m/sec. The ratio c/v = 647,948. Therefore, the equation of the spiral is: $r = 103,176\ c\theta$.

We wish to solve for the angle θ at the earth's surface, where r = 6.37x10$^6$ meters. Therefore: θ = 6.37x10$^6$/103176c = 2.05x10$^{-7}$ radians.

Recalling that 1 radian = 180/π °. Therefore, θ = 1.18x10$^{-5}$ degrees = 42.5 milli-arcseconds. This result *agrees well* with the prediction based on general relativity (ref. 11). However, the methods of general relativity are significantly more complex. The result we obtained is derived from simple considerations of the deflection of the downward accelerating virtual particle path from observer's 'A' reference frame. The calculation of the metric is not required with this method, since we are only interested in the deflection of the inertial frame.

We have seen that the angle θ represents the deflection of the downward accelerating virtual particles of the quantum vacuum with respect to the non-rotating earth. Furthermore, the deflection angle varies with height (along a spiral path). Recall that in EMQG the direction of 4D space-time curvature is the same as the direction of the virtual particle acceleration. Therefore, this angle represents the shifting of the direction of 4D space-time curvature. Now we are in a position to understand inertial frame dragging near the earth. The motion of any free falling (gravity-free) test mass is modified in the presence of the curved virtual particle path. If a test mass is in free-fall, the contribution of gravity to the motion of the mass is canceled out (at a point), and the resulting motion of the test mass is purely inertial. Therefore, the free fall is deflected by an angle of 42.5 milli-arcseconds, and the earth appears to drag the local inertial reference frame with a drag effect that varies with height.

* The next section requires a more detailed knowledge of EMQG theory, and can be bypassed without affecting our results.

## *6.    INERTIAL FRAME DRAGGING IN STRONG GRAVITY FIELDS

In general relativity inertial frame dragging is calculated using the weak gravitational field approximations to Einstein's equations. This must be done in order to overcome the legendary non-linearity of the Einstein Gravitational Field equation when the gravitational field becomes too great. In this case the mathematics becomes very difficult. However, with EQMG our physical picture at the quantum level can also explain why the strong field inertial frame dragging presents a formidable problem to the theorist.

The origin of the non-linearity for strong fields can be traced to the interaction of the outward propagating graviton particles, as they encounter the accelerated virtual fermion particles of the quantum vacuum. We have said that the graviton particle is similar to the photon particle, and therefore moves along the radius vectors of the spherical, rotating mass. This is not entirely true for strong gravitational fields. The gravitons always scatter with the virtual fermion particles of the quantum vacuum, because fermions possess 'mass-charge' (in analogy with photons, where photons scatter due to the electrical charge property of the virtual fermions).

This graviton scattering leads to a negligible correction for a small mass like the earth. However for objects with a very strong gravitational field, the scattering effect is significant. In EMQG, the non-linearity manifests itself in the way that gravitons and the quantum vacuum interact in strong gravitational fields. The graviton flux determines the motion of the surrounding virtual particles of the quantum vacuum by absorption and subsequent downward acceleration. However, the state of accelerated motion of the virtual fermion particles of the quantum vacuum in turn affect the motion of the gravitons that originate the acceleration in the first place, through the scattering process above. This makes it very difficult to calculate frame dragging under the conditions of extreme curvature and rapid rotations, such as might exist in the vicinity of black hole with EMQG

## 7. EMQG, FRAME DRAGGING, AND THE GRAVITOMAGNETIC FIELD

In EMQG, the graviton particle and the photon particle are very closely related (ref. 1 gives the properties of the graviton). Since the graviton is ultimately responsible for the gravitational field, should we not expect to find a new force field in gravitational physics that is analogous to the magnetic field of electromagnetism for moving masses? The answer to this questions turns out to be *yes*. In electromagnetism, the photon exchange process generates a magnetic field perpendicular to the direction of the moving electrical charge. Similarly, a moving mass can produce an analogous 'gravitomagnetic' force field, also directed perpendicular to the motion of the moving mass. In fact, there exists a whole gambit of equivalent electromagnetic phenomena related to magnetism that occurs with gravitational interactions such as: Gravito-Magnetic fields in general (ref. 10), Gravito-Induction, Gravito-Meissner effect, and Gravito-Lorentz force. The gravitomagnetic fields are also a consequence of Einstein's field equations, because of the close parallels between his gravitational field equation and Poisson's equation (ref. 1, section 17.4). The classical textbook on gravitation by Misner, Thorne, and Wheeler gives a good treatment on the subject.

According to general relativity (ref. 10) the weak field, slow motion limit gravitational field equations can be written in the form which emphasizes the gravitomagnetic field. When this is done, the equations for the gravitoelectric and gravitomagnetic fields become almost identical to Maxwell's equations for the electric and magnetic fields:

$\nabla \cdot \mathbf{E_g} \approx -4\pi G\rho$ , $\nabla \times \mathbf{E_g} \approx 0$ , $\nabla \cdot \mathbf{H_g} \approx 0$ , $\nabla \times \mathbf{H_g} \approx 4[-4\pi G\rho \mathbf{v}/c + (1/c)\partial \mathbf{E_g}/\partial t]$        (6.1)

where $d\mathbf{v}/dt = \mathbf{E_g} + (\mathbf{v}/c) \times \mathbf{H_g}$.

$\mathbf{E_g}$ is the gravitoelectric field and is the same as the ordinary Newtonian gravitational acceleration, $\mathbf{H_g}$ is the gravitomagnetic field which in the solar system is about $10^{12}$ times weaker than $\mathbf{E_g}$, and $\rho$ is the matter density and $\mathbf{v}$ is the velocity of the mass, and $\mathbf{E_g} = -\nabla\phi$ , and $\mathbf{H_g} = \nabla \times \gamma$ . The potentials $\phi$ and $\gamma$ are constructed in a similar manner as in electromagnetism (where the electromagnetic four-vector potential $A_\alpha$ is decomposed into an electrical scalar potential and $\psi = -A_0$ and a magnetic vector potential $A=A_j$), and are related to the space-time metric $g_{uv}$ by: $\phi \approx -0.5(g_{00} + 1)c^2$ , $\gamma_i \approx g_{0i}$. The gravitoelectric scalar potential $\phi$ is essentially the same as the time part $g_{00}$ of the space-time metric, and the gravitomagnetic vector potential **γ is** essentially the time-space part of the space-time metric.

Contrast these equations with the classical Maxwell's equations of electromagnetism, and the following differences are found:

(1) The minus sign appears in the source terms.
(2) Replacement of the charge density with mass-density multiplied by G.
(3) A factor of four in the strength of **$H_g$**.
(4) The replacement of the charge current by $G\rho$ **v**

For a rotating massive body like the earth, the gravitomagnetic field looks like the magnetic dipole field of a rotating electrically charged sphere. If one places a spinning object like a gyroscope in this field, there will be precession of the spin axis (just as a current loop in a magnetic field). This phenomena is sometimes called inertial frame dragging (we will look at an alternative way of viewing this effect in the next section).

If the existence of the gravitomagnetic fields becomes fully confirmed, we would take this as further evidence for the similarity between the photon and graviton particle exchanges. This would support the view that nature's two long-range forces (electromagnetism and gravity) operate in virtually the same way. The gravitomagnetic force component is far weaker than the observable gravitoelectric force (which corresponds to the ordinary Newtonian force of gravity), because this force is produced by pure graviton exchanges. However, the pure low-level gravitoelectric force is mixed up with the ordinary strong electrical charged component of gravity existing in all gravitational processes from the electrically charged virtual particles of the quantum vacuum. Since the electrical force from the virtual particles far exceeds the pure graviton processes, the pure gravitoelectric force component (caused by pure graviton exchanges only) is completely disguised and hard to observe. The ordinary measurable force of gravity we perceive everyday is dominated by the electromagnetic forces from the electrically charged virtual particles of the quantum vacuum, which also equalizes the fall rate of all masses.

## 8.     CONCLUSIONS

We have studied the Lense-Thirring effect (or inertial frame dragging) for a rotating mass like the earth using EMQG. The basic reason for inertial frame dragging is the finite speed of propagation of the graviton particle. This allows time for a spinning mass like the earth to rotate an observer away by a small amount while the graviton is still in flight as it propagates outwards. The finite velocity of the graviton particle along with the downward $GM/R^2$ acceleration component of the charged virtual particles of the quantum vacuum (caused by graviton exchanges between the earth and the quantum vacuum) is entirely responsible for inertial frame dragging. In EMQG, the virtual particles of the quantum vacuum determine the local frame of reference for inertia for any test mass.

The angle of deflection of the downward accelerating virtual particles of the quantum vacuum with respect to the non-rotating earth represents the amount of deflection of inertial frames, with the deflection angle varies with height according to the spiral of Archimedes. The direction of 4D space-time curvature is the same as the direction of the virtual particle acceleration. Therefore, this angle represents the shifting of the direction of 4D space-time curvature. The motion of any free falling (gravity-free) test mass is modified in the presence of the curved trajectory of virtual particles. If a test mass is in free-fall, the contribution of gravity to the motion of the mass is canceled out (at a point), and the resulting motion of the test mass is purely inertial. Therefore, the free fall frame is deflected by an angle that varies with height. We have calculated this angle at the earth's surface to be 42.5 milliarc seconds which agrees well with the result from general relativity. However, our result for frame dragging was calculated using a simple and intuitive model which illustrates the underling quantum particle physics behind inertial frame dragging.

# Figure #1 - INERTIAL FRAME DRAGGING ON THE EARTH'S EQUATOR

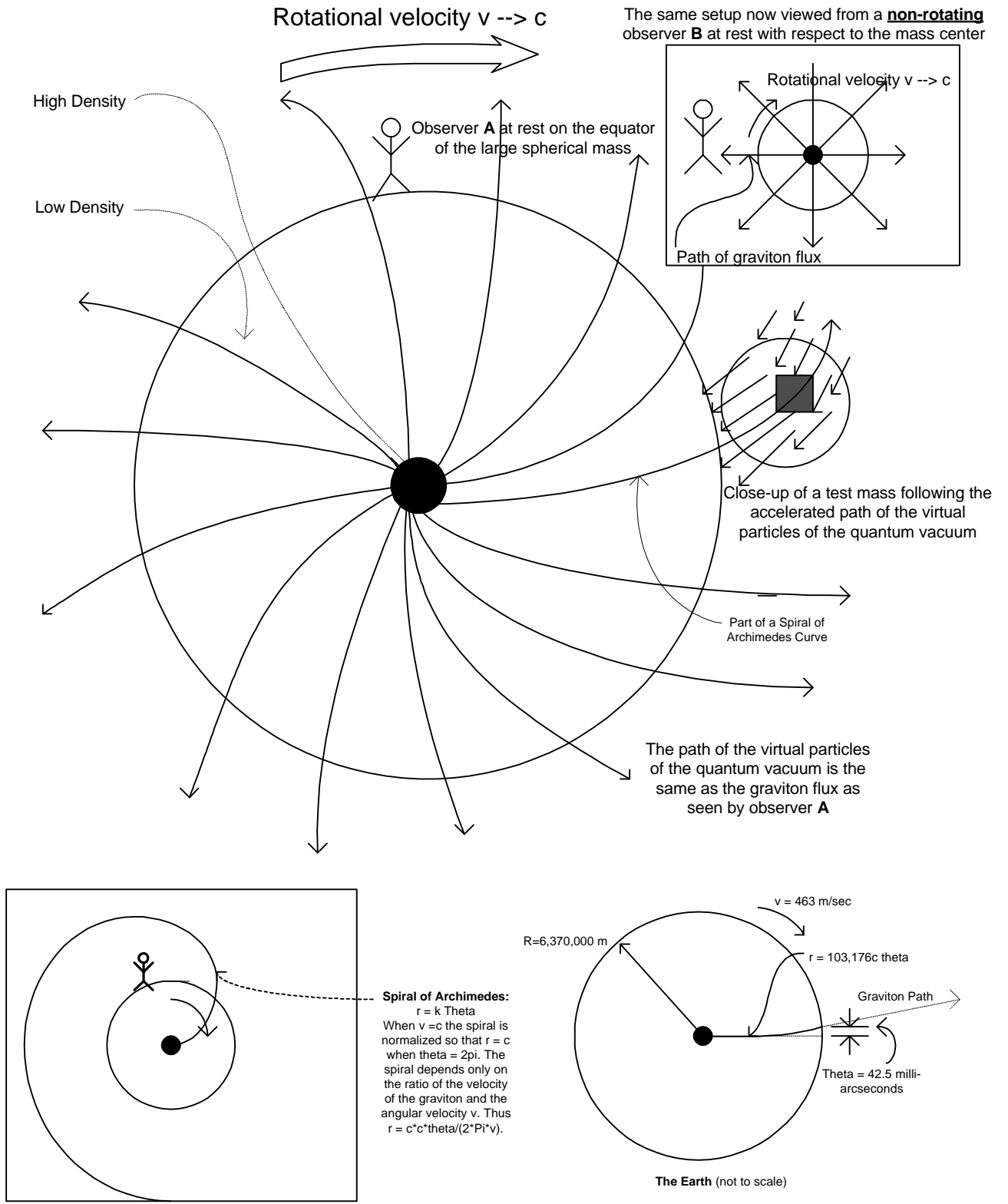